\documentclass[aps, prd, twocolumn,a4paper]{revtex4}
\usepackage{ulem}
\usepackage{color}
\usepackage{float}
\usepackage{graphicx}
\usepackage{epsfig}
\usepackage{epstopdf}
\usepackage{amsmath}
\usepackage{subfig}
\usepackage[dvipsnames]{xcolor}
\newcommand{\be}{\begin{equation}}
\newcommand{\ee}{\end{equation}}
\newcommand{\ba}{\begin{eqnarray}}
\newcommand{\ea}{\end{eqnarray}}

\usepackage{graphicx}
\usepackage{color}
\begin{document}
\title{Relaxation of baryon diffusion flow in QGP medium and its impact on thermal photon spectra}
\author{Sukanya Mitra}
\email{sukanya.mitra10@gmail.co.in}
\affiliation{National Superconducting Cyclotron Laboratory, Michigan State University, East Lansing, Michigan 48824, USA}

\begin{abstract}

In the current work the time evolution of baryon diffusion flow within the scope of second order relativistic hydrodynamic theory has 
been studied for a 2-component quark-gluon system. The relaxation time for diffusion flow is estimated using the Grad's moment method 
of many particle kinetic theory. The temperature dependence of relaxation time for diffusion flow has been obtained for different 
quark chemical potentials. Finally, the diffusion equation has been solved for a $1+1$ dimensional boost invariant system and the effect of diffusion 
flow has been studied on the thermal photon spectra in a baryon rich environment. The present analysis on the role of baryon number 
diffusion in influencing the properties of hot QCD medium could be relevant for the baryon rich matter explorations in upcoming 
experimental facilities.
{\bf PACS}: 

\end{abstract}
\maketitle

\section{Introduction}

It is now well established, that heavy ion collision experiments have been proved to be extremely efficient in describing the nuclear matter
under extreme conditions of temperature and nuclear density. The worldwide experimental facilities have produced a fortune of data over last few decades that provides strong
evidence of creating a deconfined state of fundamental partons beyond the hadrons. In this context, the results from Relativistic Heavy
Ion Collider (RHIC) at BNL \cite{STAR} and Large Hadron Collider (LHC) \cite{ALICE} at CERN are remarkable where at sufficiently high energy density 
the deconfined state has been reported to be observed given the name Quark-Gluon-Plasma (QGP), that turned out
to be more like a strongly coupled fluid. In order to describe the system properties,
so far the relativistic hydrodynamics has been proved to be an efficient theoretical approach \cite{HYDRO1}. Recently, in a number of literature
the second order causal dissipative hydrodynamics (Israel-Stewart theory) has been employed to describe the space-time evolution of the created 
fireball \cite{HYDRO2}, while the required parameters i.e. the transport coefficients (principally the shear viscosity) have been estimated using 
the underlying microscopic theory or linear response theories \cite{TRANSPORT}.

Notably, the effects of net baryon number diffusion have received much lesser attention while studying the properties of QGP within the frame work of 
dissipative hydrodynamics compared to viscosity and thermal conductivity. One of the obvious reasons is that the net baryon number
is insignificant near mid-rapidity at RHIC or LHC revealing the fact that at high energy density the collisions become transparent. However,
if the collision energy is lowered then the process of baryon stopping in the interaction zone can lead to creating mater with non-vanishing
net-baryon density. In this regard the currently running Beam-Energy Scan Program (BES) at RHIC and the NA61/SHINE experiment at CERN-SPS
as well as upcoming accelerator facility like NICA at JINR Dubna or the Compressed Baryonic Matter experiment (CBM) at the FAIR, GSI Darmstadt
are anticipated to provide the signature of a deconfined state not only by depositing high kinetic energy, but also by compressing matter at
high baryon density \cite{Friese}. Such systems are worth studying because of their rich physical properties and especially their phase
transition dynamics. At high baryon densities a first order phase transition is predicted which indicate the existence of a critical point
in order to be separated from the cross over transition of matter having vanishing net baryon density. These are the reasons of the current
interest in the study of the field of deconfined matter with finite baryon density.

There are currently a number of existing theoretical works that ventured to characterize the baryon rich systems. The net-baryon rapidity 
distribution in Au $+$ Au collisions at RHIC have been discussed in \cite{Bass}. Hydrodynamical evolution including a non-zero net-baryon 
diffusion has been studied in works like \cite{Denicol,Monnai,Floerchinger,Kapusta}. Recently the  diffusion of all conserved charges and their
coupled effects have been elaborated in \cite{Greif}. The suppression of baryon diffusion in a strongly coupled QGP has been studied in
\cite{Rougemont} using the holographic approach. In order to explore the QCD phase diagram the baryon number fluctuations have been studied
in works like \cite{Hatta,Ferreira,Plumberg} as a signal of the QCD critical point. Finally in context with this current work, the photon 
and dilepton spectra have been analyzed at FAIR and BES-RHIC energies with finite baryon density in a number of recent literature as well 
\cite{Photon-Ref}.

In this work a second order theory of baryon diffusion has been set up which gives an Israel-Stewart type time evolution equation involving
baryon diffusion flow. The microscopic framework for the required calculation is based upon multi-component covariant kinetic theory. Here,
the techniques of Grad's moment method have been employed for constructing the second order diffusion equation. The second order transport 
coefficient of this hydrodynamic equation turns out to be the relaxation time of diffusion flow. This can be understood as the characteristic 
time in which the diffusion flow, which appeared in the first place due to spatial inhomogeneity of the quark number density in a QGP medium,
decays down by wiping out those non-uniformities. The relaxation time of diffusion flow depends directly upon the first order transport
coefficient that is the diffusion coefficient itself, which has been estimated from the first order theory applying Chapman-Enskog technology.
The ratio of these two transport parameters and their temperature dependence at different quark chemical potentials are the prime aim of the current 
work. The diffusion equation has been solved for a simple 1+1 dimensional boost invariant system revealing an exponential time decay of the diffusion
flow. Finally, the out of equilibrium quark momentum distribution function has been expressed in terms of this time dependent diffusion
flow and applied for the production of thermal photon from annihilation and Compton processes. The final thermal photon yield has been compared
with the ones with no diffusion influence which reveals extremely small but quantitatively finite effects on its transverse momentum spectra.

The manuscript is organized as follows. Following the introduction, a detailed theoretical formalism has been presented in section II. This is followed by 
the result section III that includes the temperature behavior of relaxation time of diffusion flow and its influence on thermal photon spectra. 
Finally, in section IV, the work has been concluded with a brief summary and plausible outlook.

\section{Formalism} 
\subsection{Basic definitions}

We start with the basic definition of diffusion flow. In a multi-component system with $N$ number of individual particle species, the diffusion flow 
for the $k^{th}$ species is defined as,
\be
I_{k}^{\mu}=N_{k}^{\mu}-x_{k}N^{\mu}~~~~~k=1,\cdots ,N~.
\label{Eq1}
\ee
Here, the used thermodynamic quantities are discussed briefly. $N^{\mu}(x)=\sum_{k=1}^{N}N_{k}^{\mu}(x)$ is the total particle four flow which upon contracting 
with hydrodynamic four velocity $u^{\mu}(x)$ gives the total particle density $n(x)$. $N_{k}^{\mu}(x)$ is the particle four flow for the $k^{th}$ species
only, which similarly produces particle density $n_{k}(x)$ for the same. $x$ simply denotes the space-time variable. $x_{k}=\frac{n_k}{n}$ is the
particle fraction belonging to the $k^{th}$ component which owing to the relation $n=\sum_{k=1}^{N}n_k$, sums over all the species particle to give unity
$\sum_{k=1}^{N}x_k=1$.
These definitions and identities set a few constrains upon the diffusion flow mentioned in Eq.(\ref{Eq1}). First, clearly we can see that sum of diffusion
flow over all components is zero,
\be
\sum_{k=1}^{N}I_{k}^{\mu}=0~.
\label{Eq2}
\ee
Secondly, we identify the orthogonal property of diffusion flow with respect to the hydrodynamic velocity,
\be
I_{k}^{\mu}u_{\mu}=0~.
\label{Eq3}
\ee
These definitions so far are general, which means that without the particular choice of how the velocity is defined within the system.
Now the specification of velocity generates particular situations for the diffusion flow (\ref{Eq1}). 
To understand that we first decompose the particle four flow in two parts as follows,
\be
N^{\mu}=nu^{\mu}+V^{\mu}~.
\label{Eq4}
\ee
The first term on the right hand side of Eq.(\ref{Eq4}) is the part of the particle flow which purely contributed by the hydrodynamic velocity.
The remaining term $V^{\mu}=\Delta^{\mu\nu}N_{\nu}$ on right hand side is purely the irreversible part of particle four flow and can be seen
to be orthogonal to hydrodynamic velocity $u^{\mu}$ owing to the definition of the projection operator $\Delta^{\mu\nu}=g^{\mu\nu}-u^{\mu}u^{\nu}$
with $g^{\mu\nu}=(1,-1,-1,-1)$ as the metric of the system. Following Eq.(\ref{Eq1}) it can be shown that in the expression
of diffusion flow, the first term on the right hand side of Eq.(\ref{Eq4}) do not contribute where as the irreversible part $V^{\mu}$ purely
constitute the diffusion flow. So, the diffusion flow can be expressed rather as,

\be
I_{k}^{\mu}=V_{k}^{\mu}-x_{k}V^{\mu}~~~~~k=1,\cdots ,N~.
\label{Eq5}
\ee
Now in Eckart's definition, the velocity is defined as the mean particle velocity and expressed purely in terms of particle flow $N^{\mu}$,
\be
u^{\mu}=\frac{N^{\mu}}{\sqrt{N^{\nu}N_{\nu}}}~.
\label{Eq6}
\ee
Eq.(\ref{Eq6}) directly follows that the quantity $V^{\mu}=\Delta^{\mu\nu}N_{\nu}$ is zero \cite{Degroot} giving rise to a vanishing diffusion
current. Thus, in the current work throughout the Landau-Lifshitz definition of velocity has been followed where $u^{\mu}$ is expressed in terms of
momentum density and energy flow, i.e, in terms of energy momentum tensor $T^{\mu\nu}$ as followed,

\be
u^{\mu}=\frac{T^{\mu\nu}u_{\nu}}{\sqrt{u_{\rho}T^{\rho\sigma}T_{\sigma\tau}u^{\tau}}}~.
\label{Eq6A}
\ee

So far the diffusion flow has been discussed in terms of individual components of particle species. But in a thermodynamic system it is often required to
express them in term of a conserved quantum number, such as baryon number. In a general reactive mixture, the appropriate definition of diffusion flow
belonging to $a^{th}$ conserved quantum number is,
\be
I_{a}^{\mu}=\sum_{k=1}^{N}q_{ak}I_{k}^{\mu}~~~~~~~a=1,\cdots,N'~,
\label{Eq7}
\ee
with $q_{ak}$ as the $a^{th}$ conserved quantum number associated with $k^{th}$ component.

It is now time to have a microscopic definitions of the diffusion flow in terms of particle distribution.
In order to do so, let us first define the $k^{th}$ component particle four flow in terms of single particle distribution function $f_{k}$ as,

\be
N_{k}^{\mu}=\nu_k\int d\Gamma_{p_k} p^{\mu}_{k} f_k(x,p_k)~,
\label{Eq9}
\ee
with $\nu_k$ as the degeneracy. $d\Gamma_{p_k}=\frac{d^3p_k}{(2\pi)^3p_k^0}$ is the phase space factor and
$p_{k}^{\mu}$ here is the associated particle four momenta belonging to the $k^{th}$ component.
Further the particle distribution function is decomposed in an equilibrium and out of equilibrium part such as
\be
f_{k}=f_{k}^0+f_{k}^0(1\pm f_{k}^0)\phi_k~,
\label{Eq10}
\ee
with $f_{k}^0=\big[exp\{\frac{p_k^{\mu}u_{\mu}(x)-\mu_k(x)}{T(x)}\}\mp\big]^{-1}$ as the equilibrium distribution function and $\phi_k$ as the measure of deviation in particle distribution
in away from equilibrium situations.

Clearly, by these definitions it can be seen that the diffusion flow of Eq.(\ref{Eq7}) picks up only the non-equilibrium part of the distribution function leading to,
\be
I_{a}^{\mu}=\sum_{k=1}^N \nu_k (q_{ak}-x_a)\int d\Gamma_{p_k} p_{k}^{\mu} f_k^0(1\pm f_k^0)\phi_k~.
\label{Eq11}
\ee

In this work a framework giving space-time evolution of this diffusion flow has been intended. For which a second order dissipative theory
is needed to be explored. But before that a first order theory is required to be developed that gives the diffusion coefficient which is 
an important parameter of the evolution equation itself.

\subsection{First order theory}

We start with the first order relativistic transport equation of the momentum distribution functions of the constituent partons in an out of 
equilibrium system that describes the binary elastic process $p_{k}+p_{l}\rightarrow p'_{k}+p'_{l}$,
\be
p_{k}^{\mu}\partial_{\mu}f^0_{k}=-\sum_{l=1}^{N}{{\cal L}_{kl}}[\phi_{k}],~~~~~~[k=1,\cdots,N].
\label{Eq12}
\ee
The right hand side of Eq.(\ref{Eq12}) purely arises from collision process with an expression,
\ba
{\cal L}_{kl}[\phi_k]=&&\nu_{l}f_{k}^{0}\int d\Gamma_{p_l} d\Gamma_{p'_k} d\Gamma_{p'_l} f_{l}^{0}\{1 \pm f_k^{'0}\} \{1\pm f_{l}^{'0}\}\nonumber\\
&&[\phi_{k}+\phi_{l}-\phi^{'}_{k}-\phi^{'}_{l}] W(p_{k},p_{l}|p^{'}_{k},p^{'}_{l})~,
\label{Eq13} 
\ea
with $W=\frac{1}{2}(\pi)^4 \delta ^4 (p_{k}+p_{l}-p'_{k}-p'_{l})\langle|M_{k+l\rightarrow k+l}|^2\rangle$ is the interaction cross sections for the corresponding dynamical processes.
The primed quantities are nothing but the distribution functions and deviation functions with final state momenta.  
Applying Chapman-Enskog technique and ignoring terms involving viscosity the transport equation turns out to be \cite{Mitra1,Mitra2},
\begin{eqnarray}
&&p_k^{\mu}\big\{(p_k\cdot u-h_k)\big\}X_{q\mu} + p_{k}^{\mu} \sum_{a=1}^{N'-1}(q_{ak}-x_{a})X_{a\mu}=\nonumber\\&& -\frac{T}{f_k^0(1\pm f_k^0)}\sum_{l=1}^{N}{{\cal L}_{kl}}[\phi_{k}]~,
 \label{Eq14}
\end{eqnarray}
with the thermal and diffusion driving forces respectively as,
\begin{eqnarray}
X_{q\mu}=&&\frac{\nabla_{\mu}T}{T}-\frac{\nabla_{\mu}P}{nh}~,\nonumber\\
X_{k\mu}=&&[(\nabla_{\mu}\mu_{k})_{P,T}-\frac{h_{k}}{nh}\nabla_{\mu}P]~.         
\label{Eq15}
\end{eqnarray}
Here, $T$ are $P$ are the temperature and pressure of the system respectively. $\mu_k$ and $h_k$ are the chemical potential and enthalpy per particle
which are related to the total chemical potential and enthalpy per particle respectively as $\mu=\sum_{k=1}^{n}x_k \mu_k$ and $h=\sum_{k=1}^{n}x_k h_k$.
In order to be a solution of Eq.(\ref{Eq14}) $\phi_{k}$ is expressed as following,
\begin{equation}
\phi_{k}=B_{k}^{\mu}X_{q\mu}+\frac{1}{T}\sum_{a=1}^{N'-1}B_{ak}^{\mu}X_{a\mu}, 
\label{Eq16}
\end{equation}
with,
\ba
B_{k}^{\mu}=&& B_{k}(\tau_k,x)\langle p_k^{\mu}\rangle~,\nonumber\\
B_{ak}^{\mu}=&& B_{ak}(\tau_k,x)\langle p_k^{\mu}\rangle~,
\ea
where the unknown coefficients $B_{k}$'s and $B_{ak}$'s are needed to be solved from the following integral equations,
\ba
\label{Eq17A}
\sum_{k=1}^{N}{\cal L}_{kl}[B_{k}^{\mu}]=&&-\frac{1}{T}f_k^0(1\pm f_k^0) p_k^{\mu}\big\{(p_k\cdot u)-h_k\big\},\\
\sum_{k=1}^{N}{\cal L}_{kl}[B_{ak}^{\mu}]=&&-f_k^0(1\pm f_k^0) p_k^{\mu} (q_{ak}-x_a)~,
\label{Eq17B}
\ea
where $\langle p_k^{\mu}\rangle=\Delta^{\mu\nu}p_{k\nu}$ is the irreducible momentum tensor of rank $1$ and $\tau_k=\frac{p^{\mu}_k u_{\mu}}{T}$ is the $k^{th}$ particle energy 
scaled by $T$ in a comoving frame. 
Putting Eq.(\ref{Eq16}) into (\ref{Eq11}) we obtain the macroscopic proportionality relation between diffusion flow and the diffusion driving force as,
\be
I_a^{\mu}=l_{aq}X_{q}^{\mu}+\sum_{b=1}^{N'-1}l_{ab}X_{b}^{\mu}~,
\label{Eq18}
\ee
with the proportionality constants as a function of the unknown coefficients $B_{k}$ and $B_{ak}$,
\ba
\label{Eq19A}
l_{aq}=&&\sum_{k=1}^{N}\nu_k(q_{ak}-x_a)\int dF_{p_k} \langle p_k^{\mu}\rangle \langle p_{k\mu}\rangle B_{k}~,\\
l_{ab}=&&\frac{1}{T}\sum_{k=1}^{N}\nu_k(q_{ak}-x_a)\int dF_{p_k} \langle p_k^{\mu}\rangle \langle p_{k\mu}\rangle B_{bk}~,
\label{Eq19B}
\ea
with the abbreviation $dF_{p_k}=d\Gamma_{p_k}f_k^0(1\pm f_k^0)$.
Now for a purely elastic 2-component quark-gluon mixture using this identity $(\nabla^{\mu}\mu_{k})_{P,T}=\frac{T}{x_k}\nabla^{\mu}x_k$,
Eq.(\ref{Eq18}) finally provides the diffusion flow in the following manner,
\be
I_{1}^{\mu}=D_{T}\big\{\frac{\nabla^{\mu}T}{T}\big\}+nD_F\{\nabla^{\mu}x_{1}\}~,
\label{Eq20}
\ee
where,
\be
D_T=l_{1q}~,~~~~~~D_F=\frac{l_{11}T}{nx_1(1-x_1)}~,
\label{Eq20A}
\ee
are the thermal diffusion coefficient and pure diffusion coefficient respectively. The subscript $1$ in the expression of diffusion flow in Eq.(\ref{Eq20}) denotes the index of the 
conserved quantum number of interest which is the net baryon number in the present case. Now, in order to obtain  $D_T$ and $D_F$ we have to solve $B_{k}$ and $B_{ak}$ from 
Eq.(\ref{Eq17A}) and (\ref{Eq17B}) respectively. For  this purpose, the variational approximation method has been employed where $B_{k}$ and $B_{ak}$ are expressed by a polynomial
of degree $p$ as follows,
\ba
B_{k}=&&\sum_{s=0}^{p}B_{k}^{(p)s}\tau_k^s~,
\label{Eq21A}\\
B_{ak}=&&\sum_{s=0}^{p}B_{ak}^{(p)s}\tau_k^s~.
\label{Eq21B}
\ea
The coefficients $B_{k}^{(p)}$ and $B_{ak}^{(p)}$ are functions independent of particle momenta, depending only upon particle mass and thermodynamic macroscopic
quantities. Multiplying both side of Eq.(\ref{Eq17A}) and (\ref{Eq17B}) with $\nu_k \tau_k^{r}\langle p_k^{\mu}\rangle$ and integrating over $d\Gamma_{p_k}$ we
have,
\ba
\label{Eq22A}
\sum_{l,s}B_{lk}^{sr} B_{l}^{(p)s}=&&\nu_k \beta_{k}^{r}~,\\
\sum_{l,s}B_{lk}^{sr} B_{al}^{(p)s}=&&\nu_k \beta_{ak}^{r}~,
\label{Eq22B}
\ea
with,
\ba
\beta_{ak}^{r}=&&-\int dF_{pk} \langle p_k^{\mu}\rangle \langle p_{k\mu}\rangle (q_{ak}-x_a)\tau_k^r~,
\label{Eq23A}\\
\beta_{k}^{r}=&&-\frac{1}{T}\int dF_{pk} \langle p_k^{\mu}\rangle \langle p_{k\mu}\rangle \big\{(p_k\cdot u)-h_k\big\}\tau_k^r~,
\label{Eq23B}
\ea
and
\ba
B_{lk}^{sr}=&&\nu_k\nu_l\bigg[\tau_l^s  \langle p_l^{\mu}\rangle , \tau_k^r \langle p_{k\mu}\rangle\bigg] \nonumber\\
           +&&\nu_k \delta_{kl}\sum_m\nu_m\bigg[\tau_k^s  \langle p_k^{\mu}\rangle , \tau_k^r \langle p_{k\mu}\rangle\bigg]\nonumber\\
           -&&\nu_k\nu_l\bigg[\tau_l^{'s}  \langle p_l^{'\mu}\rangle , \tau_k^r \langle p_{k\mu}\rangle\bigg] \nonumber\\
           -&&\nu_k \delta_{kl}\sum_m\nu_m\bigg[\tau_k^{'s}  \langle p_k^{'\mu}\rangle , \tau_k^r \langle p_{k\mu}\rangle\bigg]~,
\label{Eq24}           
\ea
where the bracket quantity denotes,
\ba
\bigg[A,B\bigg]=&&\int d\Gamma_{p_k}d\Gamma_{p_l}d\Gamma_{p'_l}d\Gamma_{p'_l}\nonumber\\
&&f_k^0 f_l^0 (1\pm f_k^{'0})(1\pm f_l^{'0})WAB~.
\label{Eq25}
\ea
From principle of detailed balance, it can be shown that $B_{lk}^{sr}=B_{kl}^{rs}$.

Clearly, $B_{lk}^{sr}$ are the matrix elements of the collision operator with respect to various basis functions.
Now putting the expansion of $B_{k}$ and $B_{ak}$ from Eq.(\ref{Eq21A}) and (\ref{Eq21B}) in to the expression of $l_{aq}$ and $l_{ab}$ in (\ref{Eq19A}) and 
(\ref{Eq19B}), and using Eq.(\ref{Eq23A}),(\ref{Eq23B}) we have,
\ba
\label{Eq26A}
l_{aq}=&&-\sum_{k,s}\nu_k \beta_{ak}^{s} B_{k}^{(p)s}~,\\
l_{ab}=&&-\frac{1}{T}\sum_{k,s}\nu_k \beta_{ak}^{s} B_{bk}^{(p)s}~.
\label{Eq26B}
\ea

Now we need to put some constraint on both $\beta_{ak}^r$, $\beta_{k}^r$ and $B_{lk}^{sr}$ in order to estimate $l_{aq}$ and $l_{ab}$.
Here a crucial assumption is being made. The polynomial coefficients $B_{k}^{(p)s}$ and $B_{ak}^{(p)s}$ in Eq.(\ref{Eq21A}) and (\ref{Eq21B})  being independent of particle momenta,
for a mass less quark gluon system they are considered to be species independent. In such situation, by the virtue of first moment of summation
invariant the $s=0$ term in Eq.(\ref{Eq26A}), (\ref{Eq26B})  turns out to be zero, giving the lowest order approximation of $l_{aq}$ and $l_{ab}$ respectively as,
\ba
l_{aq}=&&-B^{(p)1}\sum_{k}\nu_k \beta_{ak}^{1}~,
\label{Eq27A}\\
l_{ab}=&&-\frac{1}{T}B_{b}^{(p)1}\sum_{k}\nu_k \beta_{ak}^{1} ~.
\label{Eq27B}
\ea
Now for a system with purely binary elastic collisions the summation invariant properties further constrains $\beta_{k}^0=\beta_{ak}^0=0$, which 
results in $B_{lk}^{s0}=B_{kl}^{0s}=B_{kl}^{s0}=B_{lk}^{0s}=0$ from Eq.(\ref{Eq22A}) and (\ref{Eq22B}). By virtue of these properties the lowest order approximation of the recurrence relation
(\ref{Eq22A}) and (\ref{Eq22B}) becomes,
\ba
\sum_{l,s}B_{lk}^{11} B_{l}^{(p)1}=&&\nu_k \beta_{k}^{1}~\\
\sum_{l,s}B_{lk}^{11} B_{al}^{(p)1}=&&\nu_k \beta_{ak}^{1}~.
\label{Eq28}
\ea
Applying the above recurrence relations into Eq.(\ref{Eq27A}) and (\ref{Eq27B}), for two component system the expressions of $l_{1q}$ and $l_{11}$ turn out to be,
\ba
l_{1q}=&&-\bigg\{\frac{\nu_1\beta_{11}^1+\nu_2\beta_{12}^1}{B_{11}^{11}+B_{12}^{11}}\bigg\}\nu_1 \beta_{1}^1\\
      =&&-\bigg\{\frac{\nu_1\beta_{11}^1+\nu_2\beta_{12}^1}{B_{21}^{11}+B_{22}^{11}}\bigg\}\nu_2 \beta_{2}^1~,\\
l_{11}=&&-\frac{1}{T}\bigg\{\frac{\nu_1\beta_{11}^1+\nu_2\beta_{12}^1}{B_{11}^{11}+B_{12}^{11}}\bigg\}\nu_1 \beta_{11}^1\\
      =&&-\frac{1}{T}\bigg\{\frac{\nu_1\beta_{11}^1+\nu_2\beta_{12}^1}{B_{21}^{11}+B_{22}^{11}}\bigg\}\nu_2 \beta_{12}^1~.
\label{Rq29}
\ea
 
\subsection{Second order theory}

In second order theory we intend to derive the time evolution of the diffusion flow $I_a^{\mu}$. In order to do so the technique
of Grad's moment method has been opted. In this method the deviation $\phi_k$ and so the responsible thermodynamic flows are considered to 
be small but their derivatives are not. Following this line of argument first the continuity equation and the equation of motion in
a second order theory in Landau-Lifshitz frame are given in the following manner,

\ba
Dn=&&-n\partial\cdot u-\nabla_{\mu}V^{\mu}~,
\label{Eq31A}\\
nDx_k=-&&\nabla_{\mu}I_{k}^{\mu}~,
\label{Eq31B}\\
Du^{\mu}=&&\frac{\nabla^{\mu}P}{nh}-\frac{1}{nh}\Delta^{\mu}_{\nu}\nabla_{\sigma}\Pi^{\nu\sigma}~,
\label{Eq31}
\ea
with $\Pi^{\mu\nu}$ as the viscous pressure tensor. Here, $D=u^{\mu}\partial_{\mu}$ is the covariant time derivative and $\nabla^{\mu}=\Delta^{\mu\nu}\partial_{\nu}$
is the spatial gradient.

The corresponding second order relativistic transport equation in Grad's method is as follows,
\be
\Pi_{k}^{\mu}\partial_{\mu}f^0_{k}+f_{k}^0(1\pm f_{k}^0)\Pi_{k}^{\mu}\partial_{\mu}\phi_k=-\frac{1}{T}\sum_{l=1}^{N}{{\cal L}_{kl}}[\phi_{k}]~,
\label{Eq30}
\ee
with $\Pi_k^{\mu}=\frac{p_k^{\mu}}{T}$ as the scaled particle 4-momenta.

The first term on the left hand side of Eq.(\ref{Eq30}) gives rise to a number of spatial and temporal gradients over $u^{\mu}$, $T$ and $\mu_k$. The time gradients
are eliminated by the thermodynamic identities. The terms involving time derivative over $T$ and $\mu_k$ purely contribute to the bulk viscous flow \cite{Mitra3}. 
Only the same over $u^{\mu}$ contributes to the thermal conduction and particle diffusion.  
So replacing the time derivative by the equation of motion (\ref{Eq31}) and ignoring the terms involving viscosity, the first term on the left hand side of Eq.(\ref{Eq30})
turn out to be the same thermal and diffusion driving force as before,
\ba
\Pi_{k}^{\mu}\partial_{\mu}f^0_{k}&&=f_{k}^0(1\pm f_{k}^0)\Pi_k^{\nu}\times \nonumber\\
&&\bigg\{(\tau_k-\hat{h}_k)X_{\mu} +\sum_{a=1}^{N'-1}\frac{1}{T}(q_{ak}-x_a)X_{a\mu}\bigg\}~,
\label{Eq32}
\ea
with $\hat{h}_k=\frac{h_k}{T}$.
For the remaining two terms in Eq.(\ref{Eq30}), we need to define the deviation function $\phi_k$ and its derivative. In Grad's moment method
the deviation function $\phi_k$ is expanded in particle momentum basis as the following,
\be
\phi_k=B_{k}^{\mu}(x,\tau_k)\langle\Pi_{k\mu}\rangle +\sum_{a=1}^{N'-1}\frac{1}{T}B_{ak}^{\mu}(x,\tau_k)\langle\Pi_{k\mu}\rangle~,
\label{Eq33}
\ee
with,
\be
B_{k}^{\mu}=\sum_{s=0}^{1}\{B_{k}^{s}\}^{\mu}\tau_k^s~,~~~B_{ak}^{\mu}=\sum_{s=0}^{1}\{B_{ak}^{s}\}^{\mu}\tau_k^s~. 
\label{Eq34}
\ee
The polynomial in Eq.(\ref{Eq34}) has been truncated after the first non-vanishing contribution to the diffusion flow.

Now, the zeroth-order distribution function $f_k^0$ contains a few arbitrary parameters, which are identified with the temperature, 
chemical potential, and hydrodynamic four-velocity of the system by asserting that the number density, energy
density, and hydrodynamic velocity are completely determined by the equilibrium distribution function. Following this argument
and Laudau-Lifshitz definition of velocity from Eq.(\ref{Eq6A}) which gives,
\be
\Delta^{\alpha\beta}T_{\beta\mu}u^{\mu}=0
\label{Eq35}
\ee
we find the following constraint on the expansion coefficients $\{B_{k}^s\}^{\mu}$ and $\{B_{ak}^s\}^{\mu}$,
\be
\sum_{k}\nu_k \int dF_{p_k}\langle\Pi_k^{\mu}\rangle \langle\Pi_{k\nu}\rangle\tau_k\big\{B_{k}^{\nu}+\sum_{a=1}^{N'-1}\frac{B_{ak}^{\nu}}{T}\big\}=0~.
\label{Eq36}
\ee
We define the following integral identify,
\be
\Delta^{\mu\nu}b_k^n=\int dF_{p_k}\langle\Pi_k^{\mu}\rangle \langle\Pi_{k\nu}\rangle\tau_k^n~,
\label{Eq37}
\ee
with which for a two component system Eq.(\ref{Eq36}) becomes,
\ba
\Delta^{\mu\nu}\sum_{k}\nu_k &&\bigg[\bigg\{(B_{k}^0)_{\nu}+\frac{(B_{1k}^0)_{\nu}}{T}\bigg\}b_k^1+\nonumber\\
                                   &&\bigg\{(B_{k}^1)_{\nu}+\frac{(B_{1k}^1)_{\nu}}{T}\bigg\}b_k^2\bigg]=0~.
\label{Eq38}
\ea
Now again considering a system consisting of massless particles we can assume that the polynomial coefficients $(B_k^s)^{\mu}$ and $(B^s_{ak})^{\mu}$'s of Eq.(\ref{Eq34})
are species independent and are functions of system's macroscopic thermodynamic quantities only. This directly provides the following relationship between the coefficients,
\be
\bigg\{(B^{0})_{\nu}+\frac{(B_{1}^{0})_{\nu}}{T}\bigg\}=-\bigg\{(B^{1})_{\nu}+\frac{(B_{1}^{1})_{\nu}}{T}\bigg\}  \frac{\sum_{k}\nu_k b_k^2}{\sum_{k}\nu_k b_k^1}~.
\label{Eq39}
\ee
Now putting the expression of $\phi_k$ from Eq.(\ref{Eq33}) into the expression of diffusion flow in (\ref{Eq11}),
we eventually find,
\be
\Delta^{\mu\nu}\bigg\{(B^1)_{\nu}+\frac{(B_{1}^1)_{\nu}}{T}\bigg\}=-\frac{3}{T}\frac{I_1^{\mu}}{\big\{\sum_{k}\nu_k\hat{\beta}_{1k}^1\big\}}~,
\label{Eq40}
\ee
with
\be
\hat{\beta}_{ak}^{r}=-(q_{ak}-x_a)\int dF_{pk} \langle \Pi_k^{\mu}\rangle \langle \Pi_{k\mu}\rangle \tau_k^r~.
\label{Eq41}
\ee
From (\ref{Eq39}) and (\ref{Eq40}) it readily follows,
\be
\Delta^{\mu\nu}\bigg\{(B^0)_{\nu}+\frac{(B_{1}^0)_{\nu}}{T}\bigg\}=\frac{3}{T}\frac{I_1^{\mu}}{\big\{\sum_{k}\nu_k\hat{\beta}_{1k}^1\big\}} \bigg\{\frac{\sum_{k}\nu_k b_k^2}{\sum_{k}\nu_k b_k^1}\bigg\}~.
\label{Eq42}
\ee
With Eq.(\ref{Eq40}) and (\ref{Eq42}) the deviation function $\phi_k$ of Grad's method from (\ref{Eq33}) is now completely determined
in terms of the diffusion flow and ready to apply to transport equation (\ref{Eq30}). Finally Multiplying both sides of (\ref{Eq30})
with $\nu_k\tau_k\langle\Pi_k^{\alpha}\rangle$, integrating over $d\Gamma_{p_k}$, summing over $k$ and applying derivative over $\phi_k$ we obtain,
\ba
&&X_{1}^{\alpha}\sum_{k=1}^{N}\nu_k\frac{(q_{1k}-x_1)}{T}b_k^1+\nonumber\\
&&\frac{3}{T}\frac{DI_{1}^{\alpha}}{{\sum_{k}\nu_k \hat{\beta}_{1k}^1}}\bigg[\frac{\{\sum_k\nu_kb_k^2\}^2}{\sum_k\nu_kb_k^1}-\sum_{k}\nu_kb_k^3\bigg]=\nonumber\\
&&-\frac{1}{T}\sum_{k,l}\nu_k \nu_l \int d\Gamma \tau_k \langle\Pi_k^{\alpha}\rangle\big\{\phi_k+\phi_l-\phi'_k-\phi'_l\big\}~,
\label{Eq43}
\ea
with $d\Gamma=d\Gamma_k d\Gamma_ld\Gamma'_k d\Gamma'_l f_k^0f_l^0(1\pm f_{k}^{'0})(1\pm f_{l}^{'0})W$.
Here, the thermal force term has been ignored.
The right hand side of Eq.(\ref{Eq43}) is similar to the collision integral of the first order theory apart from the fact 
that the two polynomial expansion coefficients in two techniques are different. So apart of the coefficients  $(B_{k}^{s})^{\mu}$, $(B_{ak}^{s})^{\mu}$ the 
collision integral gives back the first order transport coefficient where $(B_{ak}^{s})^{\mu}$ and $(B_{k}^{s})^{\mu}$ traces back the diffusion flow itself.

So, the final expression for the evolution of diffusion flow for a two component system at mechanical equilibrium ($\nabla_{\mu}P=0$) is,
\be
I_{1}^{\mu}=nD_F\big\{\nabla^{\mu}x_1\big\}-\tau_D D I_{1}^{\mu}~,
\label{Eq44}
\ee
where the index $1$ denotes the conserved baryon number of the system. The relaxation time of diffusion flow $\tau_D$ is,
\ba
\tau_D=9\frac{D_F}{T}\frac{{n}x_q(1-x_q)}{\{\nu_q \hat{\beta}_{1q}^{1}\}^2} \bigg[ \frac{\{\nu_q b_q^2\}^2}{\nu_q b_q^1}-\nu_q b_q^3 \bigg]~.
\label{Eq45}
\ea
The subscript $q$ denotes the quark components. Here another assumption is made, that is when only baryon diffusion flow is being concerned, the gluonic 
contribution in $\phi_k$ is negligible compared to quark contribution.
In comparison with the first order theory given in Eq.(\ref{Eq20}) we identify the addition of the second term on the right hand side of (\ref{Eq44})
which gives the time evolution of $I_{1}^{\mu}$ preserving the causality of the theory.

\subsection{Photon spectra}
In the previous section we realized that the deviation function $\phi_k$ from Eq.(\ref{Eq33}) as well as the out of the equilibrium distribution function
of the quarks in a quark-gluon-plasma system can be expressed by the baryon diffusion current in it. For this purpose we need to solve $I_1^{\mu}$ as a
function of space-time variable from Eq.(\ref{Eq44}) with the help of Eq.(\ref{Eq31A}) and (\ref{Eq31B}). For this purpose the equations are first
expressed for a system with $1+1$ dimensional expansion in the $z$ direction. The concerned space-time variables are chosen to be as the proper time $\tau$ and space-time 
rapidity $\eta$ which are related to the original variables ($x^{\mu}=t,z$) as $t=\tau \textrm{cosh}\eta$ and $z=\tau \textrm{sinh}\eta$.
Further the velocity and diffusion flow are defined as follows,
\ba
u^{\mu}=&&(\textrm{cosh}\eta,0,0,\textrm{sinh} \eta)~,\\
I_{B}^{\mu}=&&I_{B}(\textrm{sinh}\eta,0,0,\textrm{cosh} \eta)~,
\label{Eq47}
\ea
such that $u_{\mu}I_B^{\mu}=0$. The subscript $B$ is used from now on instead of $1$ to express the baryon number.

With help of this set up the hydrodynamical equations finally turn out to be,

\ba
\tau_D \frac{\partial I_B}{\partial \tau}=&&nD_{F}\frac{1}{\tau}\frac{\partial x_B}{\partial \eta}-I_B~,
\label{Eq48A}\\
\frac{\partial n}{\partial \tau}=&&-\frac{n}{\tau}-\frac{1}{\tau}\frac{\partial V}{\partial \eta}~,
\label{Eq48B}\\
\frac{\partial x_B}{\partial\tau}=&&-\frac{1}{n}\frac{1}{\tau}\frac{\partial I_B}{\partial \eta}~.
\label{Eq48}
\ea

From the above set of equations for a boost invariant system the solution of diffusion flow comes out to be,
\be
I_{B}(\tau)=I_B^0 exp\{-\tau/\tau_D\}~,
\label{Eq49}
\ee
with $I_B^0$ as the initial value of the diffusion flow.
Now, to construct the out of equilibrium distribution function the momenta of the interacting particles are defined as,
\be
p_k^{\mu}=(m_{k\bot}\textrm{cosh}y_k,p_{kT}\textrm{cos}\phi_k,p_{kT}\textrm{sin}\phi_k,m_{k\bot}\textrm{sinh}y_k)~,
\label{Eq50}
\ee
with $m_{k\bot}=\sqrt{p_{k\bot}^2+m_k^2}$ as the particle transverse mass, $m_k$ as particle mass, $p_{k\bot}$ as transverse momenta
and $\phi_k$ as the angle of particle's momenta along z-axis. With these prescriptions the out of equilibrium distribution function
comes out to be,
\ba
\delta f_k=&&f_k^0(1\pm f_k^0) \phi_k\nonumber\\
          =&&f_k^0(1\pm f_k^0) \frac{1}{\nu_k\beta_{1k}^1}I_{B}^0exp(-\tau/\tau_D)\times\nonumber\\
          &&\bigg\{\frac{m_{k\bot}^2}{T}\textrm{sinh}(y_k-\eta)\textrm{cosh}(y_k-\eta)\nonumber\\
          &&-\frac{b_k^2}{b_k^1}m_{k\bot}\textrm{sinh}(y_k-\eta)\bigg\}~.
\label{Eq50A}         
\ea
So using the total particle distribution function as $f_k=f_k^0+\delta f_k$, the transverse momentum distribution of thermal photons can be given by \cite{Mitra4},  

\begin{eqnarray}
\frac{dR}{d^2p_Tdy}=\frac{\mathcal{N}}{16(2\pi)^8}
\int p_{1T}dp_{1T}dp_{2T}d\phi_1dy_1dy_2\nonumber\\
f_1 f_2 (1\pm f_3)\times \frac{\overline{\arrowvert \mathcal{M}\arrowvert^2}}
{\arrowvert p_{1T}\sin(\phi_1-\phi_2)+
p_{T}\sin\phi_2 \arrowvert_{\phi_2=\phi_2^0}}~.
\label{Eq51}
\end{eqnarray}
$\cal{N}$ is the overall degeneracy for the reaction under consideration and $\overline{|{\mathcal {M}}|^2}$ is the square of the invariant 
amplitude for the QGP interactions, which in the present case have been taken to be the annihilation ($q\bar{q}\rightarrow g\gamma$) and
Compton ($qg\rightarrow q\gamma$) processes.
So the  photon yield ($dN/d^2p_Tdy$) at mid-rapidity obtained after performing a space-time integration over the evolution history is given 
as the following,
\begin{equation}
\frac{dN}{d^2p_Tdy}\mid_{y=0}=
\int d^4x\left[\frac{dR}{d^2p_Tdy}\mid_{y=0}\right]~,
\label{Eq52}
\end{equation}
which captures the effect of baryon diffusion flow within the medium via the out of equilibrium quark distribution functions. 

Now there a question arises regarding the initial value of baryon diffusion flow $I_B^0$. From Eq.(\ref{Eq48A}) it can be observed that the Navier-Stokes 
approximation can predict an initial value $I_B^0 \sim T_i^2/\tau_i$, $T_i$ and $\tau_i$ being the initial value of temperature and time. 

\section{Results}

From the above analysis the behavior of the estimated quantities is presented in this section for a mass less QGP system with number quark of flavors $N_f=3$.

\begin{figure}[ht]
\includegraphics[scale=0.35]{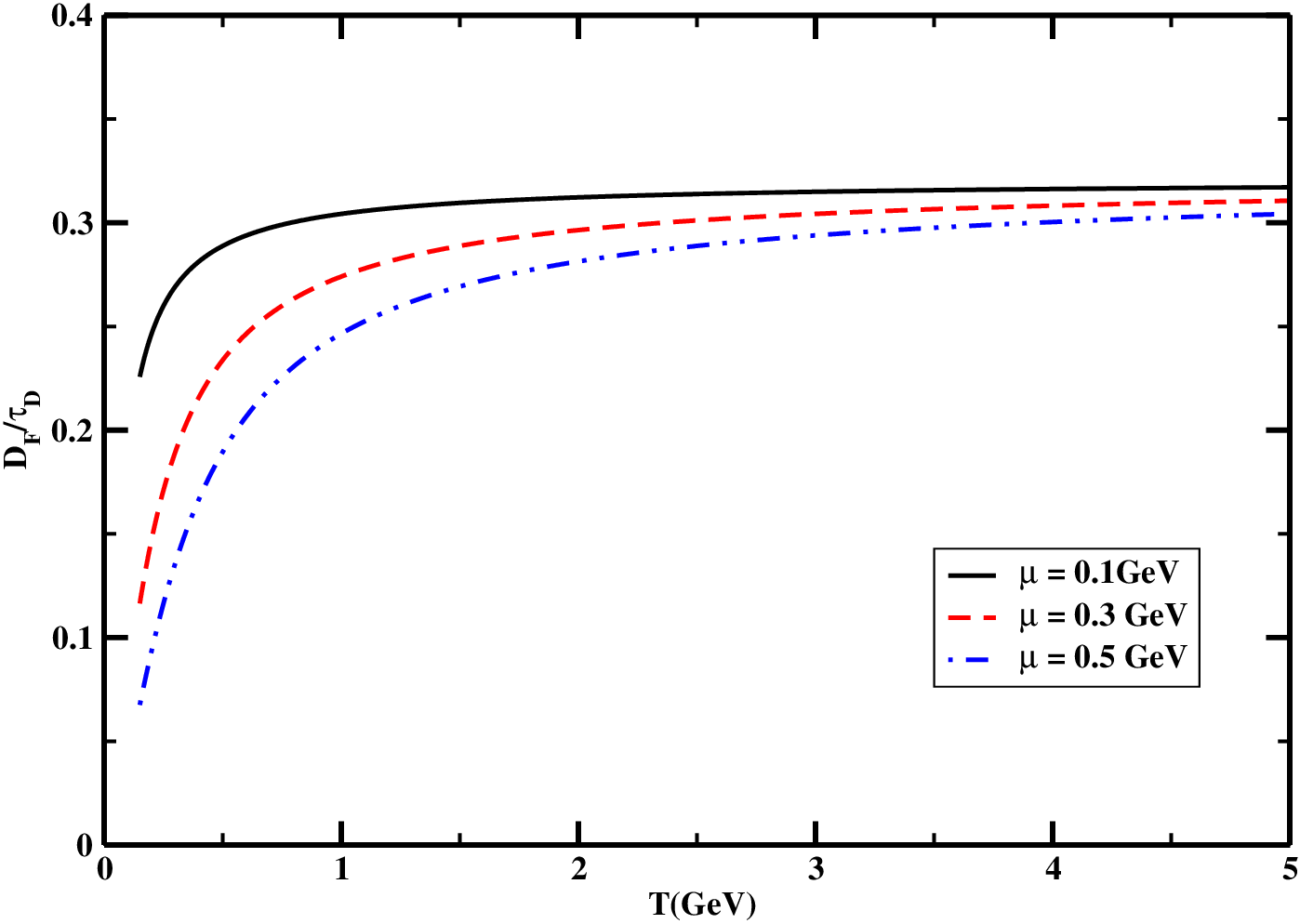}
\caption{Ratio of diffusion coefficient to relaxation time of diffusion flow as a function of temperature at different quark chemical potentials.}
\label{Dbytau}
\end{figure}

Following the general definition of Fick's law of diffusion, Eq.(\ref{Eq44}) gives the Fick's second law where Eq.(\ref{Eq20}) gives the first law ignoring 
the temperature gradient \cite{Kelly,Bhalerao,Gupta}. Before providing the temperature dependence of $\tau_D$, the dimensionless ratio $D_F/\tau_D$ between 
the first and second order transport coefficients for diffusion process (which is clearly the right hand side of Eq.(\ref{Eq45}) except $D_F$) has been plotted 
as a function of temperature in Fig.(\ref{Dbytau}) for three sets of quark chemical potentials $\mu_q$. Ref.\cite{Kelly,Bhalerao} have predicted this ratio to be 
equal to the speed of sound $c_s^2$ within the medium. For temperature beyond $1$ GeV the ratio $D_F/\tau_D$ approaches the Stefan-Boltzmann limit of this value (around $3.1$) which
is the consistent sound speed in high temperature QGP. However, in the temperature region $0.15-1$ GeV the ratio shows strong temperature dependence ranging
between $2-3$. This can be considered as a prediction of non-trivial behavior of thermodynamic parameters around transition temperature $T_c$.
\\
\\

\begin{figure}[ht]
\includegraphics[scale=0.35]{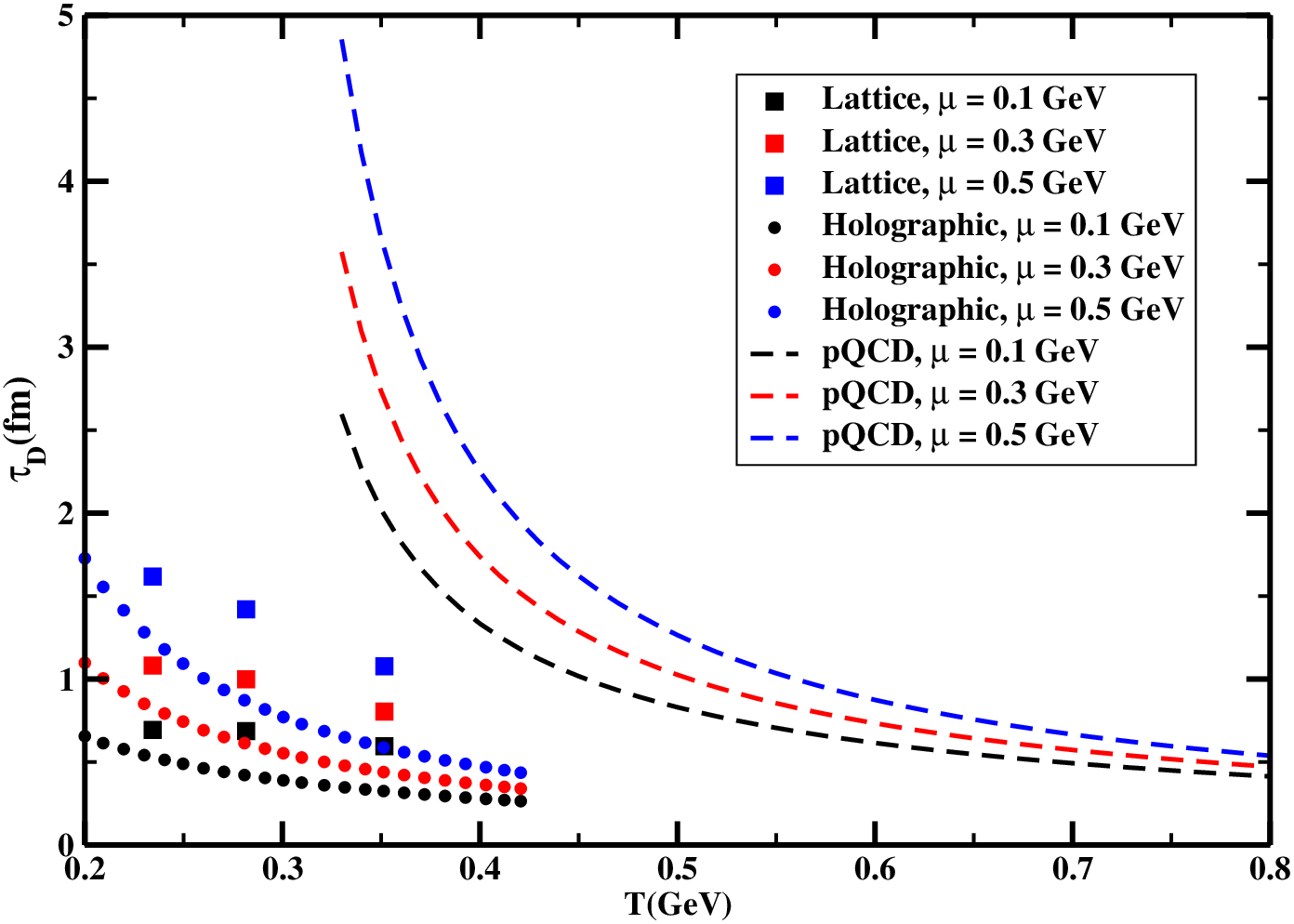}
\caption{Relaxation time of diffusion flow as a function of temperature with varied $\mu_q$ using diffusion coefficient from different approaches.}
\label{tauD}
\end{figure}

In Fig.(\ref{tauD}) $\tau_D$ has been plotted as a function of temperature for varied values of $\mu_q$. In order to estimate $\tau_D$ three different values for diffusion 
coefficient $D_F$ have been taken, the lattice values from \cite{Aarts}, the holographic values from \cite{Finazzo} and pQCD values from \cite{Mitra5}. In Fig.(\ref{tauD}) 
$\tau_D$ is showing larger values at higher $\mu_q$. It simply indicates that with high initial baryon chemical potential the diffusion takes longer time to decay down. 
However, the ADS/CFT prediction of $\tau_D\sim 1/T$ \cite{Natsuume}, which have been followed in hydrodynamic simulations like \cite{Denicol,Plumberg}, can be 
directly traced back from the nature of $D_F$ itself which is inversely proportional to temperature. At small $\mu_q$ the values of $\tau_D$ for pQCD case is $\sim 1fm$, where as
for holographic scenario they are quite small $\sim 0.1 fm$, which are consistent with the prediction by \cite{Bhalerao}. 
\\
\\

\begin{figure}[ht]
\includegraphics[scale=0.35]{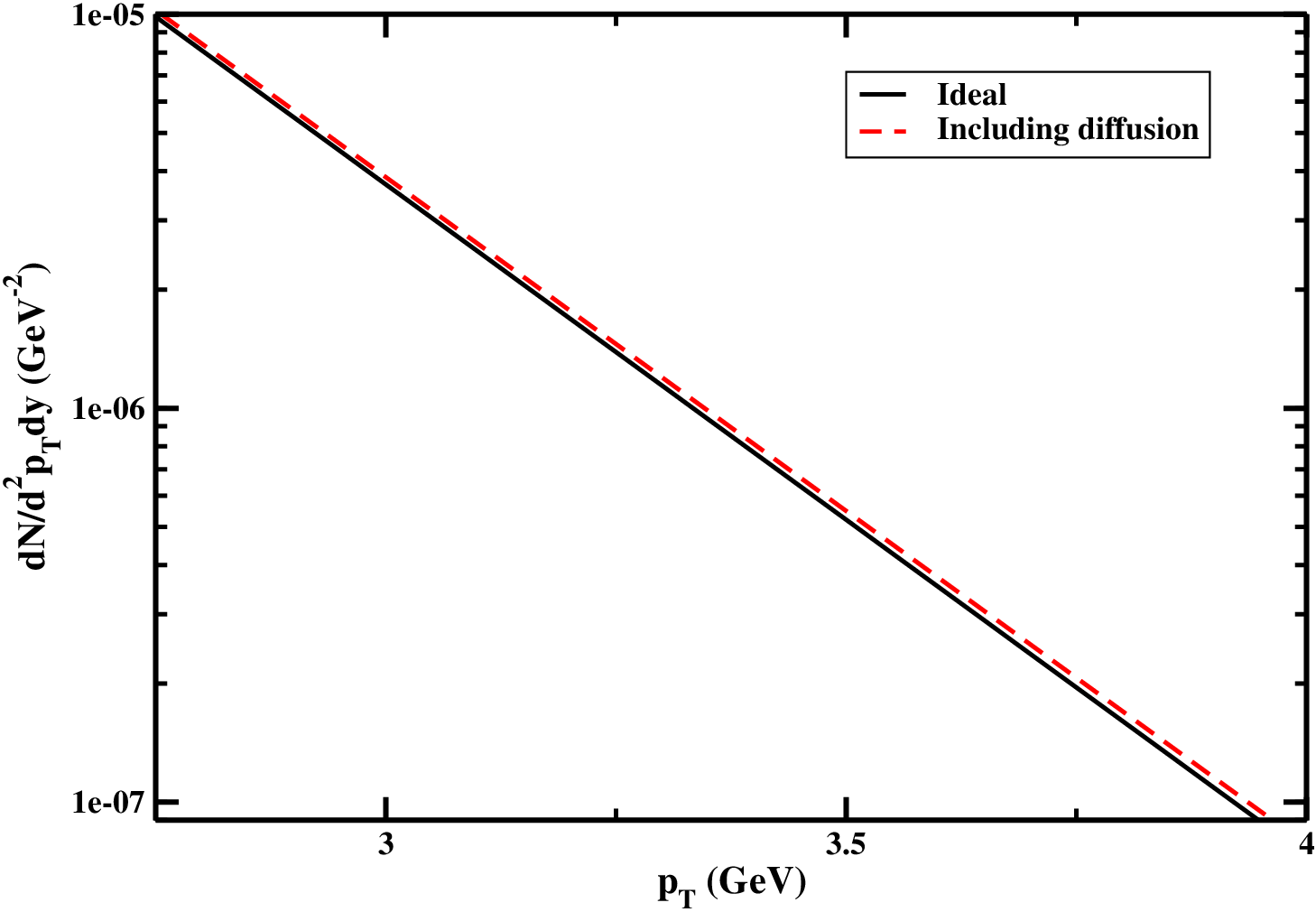}
\caption{Transverse momentum spectra of thermal photons with and without including baryon diffusion flow.}
\label{Photon} 
\end{figure}

Fig.(\ref{Photon}) displays transverse momentum spectra of thermal photons from process $q\bar{q}\rightarrow g\gamma$ and $qg\rightarrow q\gamma$. The solid line of the 
plot do not consider any diffusion flows, where as the dashed line indicates that the out of equilibrium distribution function from Eq.(\ref{Eq50A}) which involves
diffusion flow, has been incorporated in the quark momentum distribution of the collision processes. The expansion dynamics follows $n\tau=$ constant for a boost-invariant
system from Eq.(\ref{Eq48B}). The values of transition temperature ($T_c$) and and the corresponding baryon chemical potential ($\mu_c$) has been taken to be,
$T_c=0.16$ GeV and $\mu_c=0.357$ GeV from \cite{Dutta} where they have been predicted by fixing the entropy per baryon ratio ($S/B$). The initial values of time,
temperature and $\mu_q$ have been fixed at $\tau_i=0.3$ fm, $T_i=0.3$ GeV and $\mu_i=0.67$ GeV respectively. The observed effect of baryon diffusion on thermal photon
spectra is significantly small. In the $p_T$ range $2.75-4$ GeV the small but quantitative effect is visible which is slightly increasing with increasing $p_T$.
Hence the finite baryon flow effect on photon $p_T$ spectra is small but still observable.
\\
\\

\begin{figure}[ht]
\includegraphics[scale=0.35]{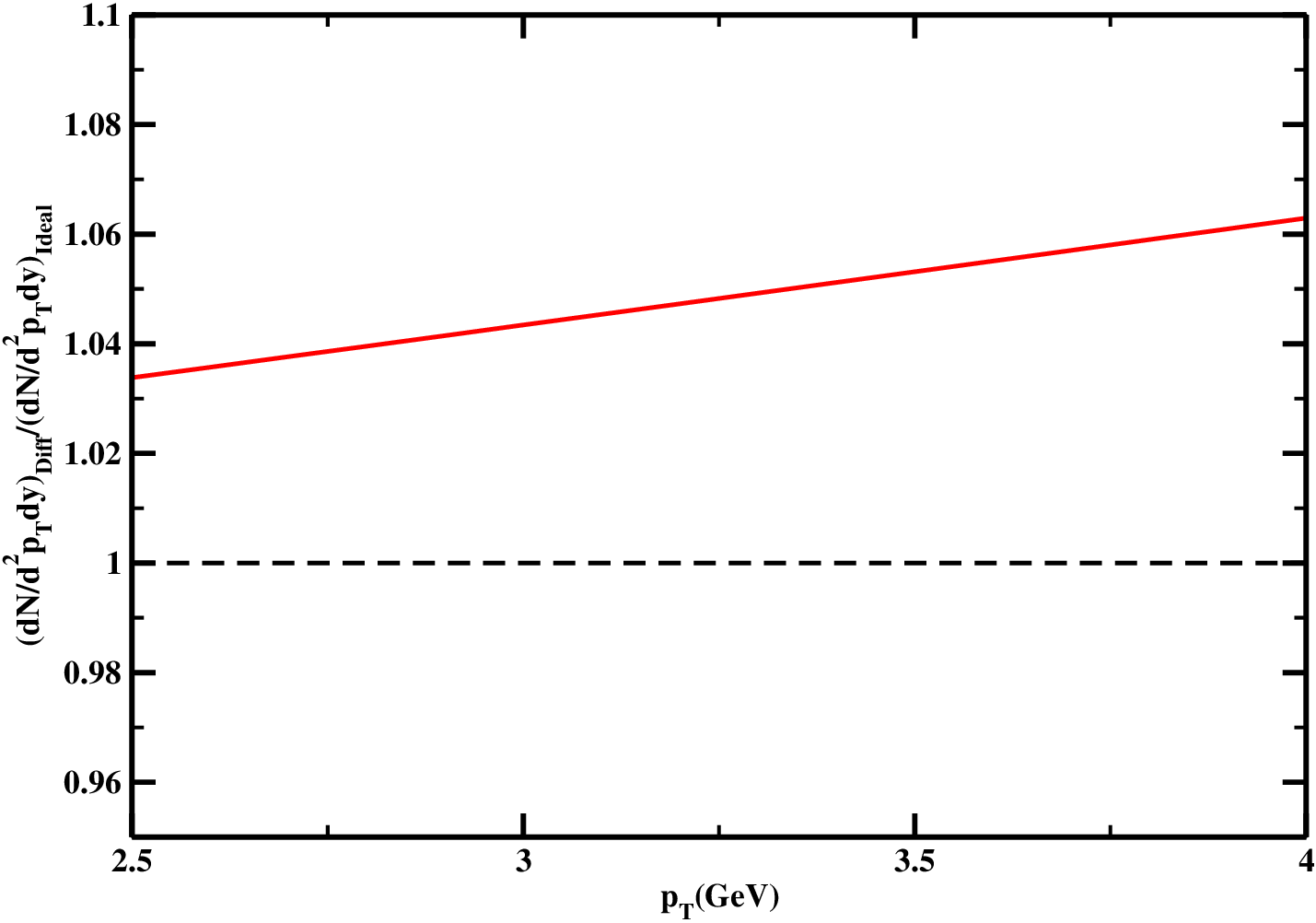}
\caption{Ratio of thermal photon spectra with and without baryon diffusion flow.}
\label{Photon-ratio} 
\end{figure}

To observe this quantitative increment in thermal photon spectra, the ratio between $dN/d^2p_Tdy$ with and without baryon diffusion effects has been plotted as a function 
of $p_T$ in Fig.(\ref{Photon-ratio}). The solid line shows that at $p_T\sim 2.5$ GeV the increment is around $3\%$ which reaches up to $6\%$ at $p_T\sim 4$ GeV.
\\
\\

\section{Conclusion and Outlook}
In the current work the baryon diffusion flow of a QGP system with finite quark chemical potential has been studied using the Grad's moment method of 
relativistic kinetic theory. The second order hydro equations associated with quark fraction and diffusion flow have been set up which gives the diffusion 
relaxation time scale as a function of $T$ and $\mu_q$. The involved microscopic theory modifies the quark distribution functions in the due course reflecting the 
effects of diffusion flow on their momentum distribution. Finally, the transverse momentum spectra of thermal photons from those quarks have been estimated and 
compared with the ideal case that does not include diffusion for a $1+1$ dimensional boost invariant system. The estimated values of $D_F/\tau_D$ ratio and $\tau_D$ itself 
come up to be consistent with ADS/CFT or other microscopic models. The incorporation of diffusion flow in quark distribution produces small but finite effect on 
thermal photon spectra which shows $3-6\%$ increment in the $p_T$ range $2.5-4$ GeV.

Immediate future extensions of the work would include generalization of the baryon diffusion flow for (2+1)-dimensional and (3+1)-dimensional hot QCD/QGP systems. 
In addition, explorations on the impact of baryon richness on the traditional first and second order transport coefficients such as shear and bulk viscosities, thermal 
conductivity and electrical conductivity could be another related direction to focus on in the near future.
 
In the view of the current scenario and expected data from upcoming accelerator experiments, in depth study of baryon rich systems and identifying their
physical properties are turning out to be one of the most promising frontiers of heavy ion physics. The open problems like the equation of state of such matter
and the applicability of the existing models, the nature of phase transition from hadronic to partonic matter and the search for a critical point that can be manifested 
in event-by-event fluctuation of conserved net baryon number, are the open horizon to step forward. Thus, the detailed investigation of relevant observables with critical
analysis and proper theoretical set up in this direction is likely to be ventured in the upcoming days making the high density heavy ion collision a most promising
field of study.

\acknowledgments
I duly acknowledge the Science and Engineering Research Board (SERB) and Indo-U.S. Science and Technology Forum (IUSSTF) for funding the Indo-US postdoctoral fellowship 
and providing the financial support. I also thank Prof. Scott Pratt and Prof. Vinod Chandra for useful discussions and helpful remarks.

\end{document}